\documentclass[11pt]{article}
\usepackage{graphicx}
\usepackage{placeins}
\usepackage{float}
\usepackage{subfigure}
\usepackage[toc,page]{appendix}

\usepackage{amssymb}
\usepackage{epsfig}
\begin{document}
\title{{\bf{\Large Thermodynamics of Sultana-Dyer Black Hole}}}
\author{ 
 {\bf {\normalsize Bibhas Ranjan Majhi}$
$\thanks{E-mail: bibhas.majhi@mail.huji.ac.il}}\\ 
{\normalsize Racah Institute of Physics, Hebrew University of Jerusalem,}
\\{\normalsize Givat Ram, Jerusalem 91904, Israel}
\\[0.3cm]
}

\maketitle

\begin{abstract}
   The thermodynamical entities on the dynamical horizon are not naturally defined like the usual static cases. 
Here I find the temperature, Smarr formula and the first law of thermodynamics for the Sultana-Dyer metric which is 
related to the Schwarzschild spacetime by a time dependent conformal factor. To find the temperature ($T$), the chiral anomaly 
expressions for the two dimensional spacetime are used. This shows an application of the anomaly method to study 
Hawking effect for a dynamical situation. Moreover, the analysis singles out one expression for temperature among two existing 
expressions in the literature. Interestingly, the present form satisfies the first law of thermodynamics. Also, it relates the Misner-Sharp energy ($\bar{E}$) and the horizon entropy ($\bar{S}$) by an algebraic expression $\bar{E}=2\bar{S}T$ which is the general form of the Smarr formula. This fact is similar to the usual static black hole cases in Einstein's gravity where the energy is identified as the Komar conserved quantity. 
\end{abstract}

\section{Introduction and Motivation}
For more than thirty years, black hole physics has been one 
of the most active areas of research in both classical and 
quantum gravity. However, almost all the research has been 
limited to stationary black holes, represented in general by the 
Kerr-Newman solution. All these exact solutions share two basic 
characteristics. The first one is asymptotic flatness, which 
means that spacetime is Minkowskian at large distances from 
the black hole. The second one is time-independence, which 
implies the existence of a timelike Killing vector field outside 
the horizon. This generates the Killing horizon, used 
to describe the event horizon, in these black hole spacetimes. 
These black hole solutions, which are considered as good 
approximations to real astrophysical black holes, have been very 
useful for studying physical effects such as the classical 
tests and quantum black hole evaporation.

However, in a more realistic situation, the black hole is part 
of a cosmological model and it may be surrounded by a local 
mass distribution. In this case, the spacetime becomes 
cosmological and non-flat at large spatial distances from the 
black hole, and it may also cease to be time-independent. One 
expects significant differences in the structure and properties 
of these so-called cosmological black holes from the well known 
stationary ones, where even the passage from
the Schwarzschild to the Kerr solution brings profound changes.

My goal in this paper is to give a thorough study of  the 
thermodynamics of Sultana-Dyer (SD) black hole \cite{Sultana:2005tp}.
This is a inhomogeneous and time dependent solution of general theory of 
relativity (GR) in presence of two noninteracting perfect fluids. One fluid 
is a timelike dust and other one is a null dust. The spacetime is 
asymptotically Friedmann-Lemaitre-Robertson-Walker (FLRW). It is 
conformal to the Schwarzschild black hole where the conformal 
factor is time dependent. Hence the spacetime metric is evolving 
with time and so the usual way of defining the thermodynamical entities 
on the horizon does not work in a straightforward way. 
The primary interest here must be on the expressions for temperature, 
entropy,  and energy. These quantities are not so obviously 
defined because the usual stationary formulation can not be 
applied  without modifications. Hence, the law of thermodynamics,  
the Smarr formula, {\em  etc.} have yet to be introduced in a 
consistent way. 

The main concern is to find the temperature. In the 
literature,  there exist two forms of temperature for the SD 
spacetime: one  is  independent of time \cite{Sultana:2005tp,Faraoni:2007gq}, 
while another  depends on  time \cite{Saida:2007ru}. But which  
one is correct must be sorted out in order to develop the full 
thermodynamics. This discrepancy can be avoided by calculating 
the emission spectrum from  the dynamical horizon, from which it 
is easy to read off the expression for the temperature.  This is 
actually inspired by Hawking's original calculation  for a 
collapsing dust in the Schwarzschild background \cite{Hawking:1974sw}. But 
a straightforward application of Hawking's procedure is not  
feasible here because in the original calculation  the 
radiation spectrum was computed  by evaluating the Bogoliubov 
coefficients relating ingoing and outgoing modes of positive and 
negative frequencies,  which are defined by familiar boundary 
conditions when the spacetime is Minkowskian at infinity. In the 
present case, instead, the spacetime is asymptotically  
FLRW and it is obtained by 
means of a conformal transformation of the Schwarzschild metric 
under which the Bogoliubov coefficients are not invariant. 

To overcome this difficulty, we will take the advantage of the fact that 
the SD metric is conformal to Schwarzschild (SC) black hole.
Now since SC is static, its Hawking radiation spectrum is well known. 
Therefore to find the same for the conformally connected metric, one way, 
may be, is to obtain the relations among the required quantities in two spacetimes.
Here this type of logic will be adopted.
We will obtain the radiation spectrum by 
using the two dimensional gravitational (chiral) anomaly expressions. Normally, this 
method was used for stationary black hole spacetimes. We are 
interested  to use this method not only for finding the 
temperature, but also to study the applicability of this anomaly 
method to a dynamical spacetime,  which is unknown in the 
literature.  In this method one has to find the near-horizon effective 
theory since the anomaly expressions are valid for near-horizon 
theory, which is based on a two-dimensional ($t-r$) metric for 
black holes. Once we  see that for our case 
the near-horizon theory is again two-dimensional, then it will be 
easy to find the flux from the anomaly  with 
the proper modifications.

The steps are as follows. It will be observed that the near horizon 
SD metric is effectively two dimensional. So the two dimensional
 anomaly expressions 
are identical in form for both Schwarzschild and SD spacetimes. This will help us to find the transformation rule of the two dimensional anomalous energy-momentum 
tensor under the conformal transformation. It will present a connection 
between the components of the stress-tensor for both spacetimes.  
As already known that the anomaly equations are easy to solve for 
static cases to obtain the stress-tensor, the above connection will lead to the required component to find the flux from the SD horizon. Then the expression for the 
temperature will be identified. We shall also show that the temperature is 
identical to the non-affinity parameter corresponding to a null vector which 
is conformal to that of the Schwarzschild spacetime. The horizon for the present analysis will be shown as a trapping one.

Next the Misner-sharp energy ($\bar{E}$), defined in \cite{Misner:1964je}, 
will be evaluated by performing the proper transformation of it. The entropy 
($\bar{S}$) will be found out using the usual area law. Having obtained 
all these, we shall show that they are related by the relation $\bar{E}
=2\bar{S}T$ where the $T$ is the temperature of the SD horizon. A similar 
expression was also observed earlier for the static and stationary cases 
in GR \cite{Padmanabhan:2003pk,Banerjee:2010yd,Banerjee:2010ye} and it has been shown that such a relation is actually the 
usual Smarr formula \cite{Smarr:1972kt}. For the present case we will give the Smarr 
formula using the above result. Finally, it will be shown that the form of 
the temperature, obtained in this paper, satisfies the first law of 
thermodynamics where the energy is the Misner-sharp energy.

Let us now summarise how the paper will be organized. In section \ref{metric}, 
a small introduction of the SD metric will be presented. Next I will give the 
derivation of the temperature from the two dimensional chiral gravitational 
anomaly expressions. Section \ref{Smarr} will contain the expressions for the 
Smarr formula and the first law of thermodynamics. Final section will be devoted 
for the conclusions. An appendix on the dimensional reduction for SD metric 
will also be given at the end.

\section{\label{metric}SD spacetime and null coordinates: a brief review}
Here we shall present a summary of the SD metric without much details. This will serve the main purpose of the paper. Also the usual null coordinates will be introduced which will turn out to be the convenient coordinate system for the present calculations. The near horizon two dimensional effective metric will be expressed in this frame.
 
   The SD metric is a solution of GR with two noninteracting perfect fluids: 
one is timelike and the other one is null-like. It is asymptotically FLRW and can be shown to be conformal to the Schwarzschild black hole. For details, see 
\cite{Sultana:2005tp}. The conformal factor being time dependent, the spacetime 
is an evolving one.
The form of the line element is of the following form \cite{Sultana:2005tp}:
\begin{equation}
ds^2 = a^2(\eta)\Big[-d\eta^2+dr^2+r^2(d\theta^2+\sin^2\theta d\phi^2)+\frac{2m}{r}(d\eta+dr)^2\Big]~,
\label{SD1} 
\end{equation}
where $m$ is a positive constant, identified as the mass of the Schwarzschild black hole. $\eta$, $r$ are the time and the radial coordinates, 
respectively. The conformal factor is given by $a(\eta) = \eta^2$. Interestingly, this metric is conformal to Schwarzschild metric. 
To see this, let us perform the coordinate transformation $\eta = t + 2m \ln(r/2m - 1)$ which turns Eq. (\ref{SD1}) into
\begin{equation}
ds^2 =  a^2(t,r)\Big[-\Big(1-\frac{2m}{r}\Big)dt^2 + \frac{dr^2}{\Big(1-\frac{2m}{r}\Big)}+ r^2(d\theta^2
+\sin^2\theta d\phi^2)\Big]~.
\label{SDSC}
\end{equation}
Using the coordinate relation between the conformal time and the Schwarzschild time, it is easy to see that the conformal 
factor is given by \cite{Faraoni:2013aba},
\begin{equation}
a(t,r) = \Big(t+2m\ln\Big|\frac{r}{2m}-1\Big|\Big)^2~.
\label{a}
\end{equation}
In the subsequent analysis, we shall associate the thermodynamical quantities on the horizon $r=2m$ which will be shown to be the 
trapping horizon of SD metric.

For our future purpose, let us now introduce the usual null coordinates:
\begin{equation}
u = t-r_*; \,\,\ v=t+r_*~;
\label{null} 
\end{equation}
where the tortoise coordinate is given by $dr_* = dr/F(r)$ with $F(r) = 1-2m/r$. 
It is well known that for the Schwarzschild spacetime, near the horizon the physics is a conformal theory which is 
effectively governed by the ($t-r$)-sector of the full metric (For details, see \cite{Majhi:2011yi}). This two dimensional sector 
in null coordinates takes the following form:
\begin{equation}
ds^2_{\textrm{Sch}} = -\frac{F(r)}{2}(dudv+dvdu)~.
\label{2SCH}  
\end{equation}
The above conclusion is basically confined to the static or stationary spacetimes. For the SD metric, it is also 
possible to show that near the horizon, the effective theory is again conformal and driven by the ($t-r$)-sector 
of the metric (\ref{SDSC}). Expressing this in null coordinates we obtain
\begin{equation}
ds^2_{\textrm{eff}} = a^2(t,r)\Big[-\frac{F(r)}{2}(dudv+dvdu)\Big]~.
\label{2SD}
\end{equation}
For completeness, a detailed dimensional reduction for the SD metric in the near horizon limit has been presented in the 
Appendix. Here note that the effective spacetimes are again related by the same time dependent conformal factor. In the next section the above two spacetimes will be used to find the radiation flux and the temperature of the SD horizon.

\section{\label{temp}Gravitational anomalies and Hawking temperature}
    The gravitational theories are usually accompanied by gravitational anomalies at the quantum level 
\cite{AlvarezGaume:1983ig,Polyakov:1981rd,Leutwyler:1984nd}. In two dimensional case, 
either trace or both trace and diffeomorphism anomalies appear depending on the symmetries of the theories. In non-chiral case 
(i.e. vector theory) one retains the diffeomorphism symmetry at the cost of conformal symmetry. So only trace anomaly appears. 
The Hawking radiation has been studied using such anomaly long ago \cite{Christensen:1977jc}. On the contrary, in the chiral case, both the trace and diffeomorphism 
anomalies appear. Use of these to study Hawking effect has been done in \cite{Robinson:2005pd,Banerjee:2007qs}. Almost all were confined for the static cases. Here I will use these anomalies to find the emission spectrum for the SD metric which is a dynamical one. For chiral case, the explicit expressions are given by \cite{AlvarezGaume:1983ig,Leutwyler:1984nd},
\begin{eqnarray}
T\equiv g^{ab}T_{ab} = \frac{R}{48\pi}; \,\,\,\,\  \nabla_a T^{ab}=\frac{1}{96\pi}\tilde{\epsilon}^{bc}\nabla_c R
\label{2.01} 
\end{eqnarray}
where $R$ is the two dimensional Ricci scaler corresponding to the ($t-r$)-sector of the metric and $\tilde{\epsilon}^{ab} 
= \epsilon^{ab}/\sqrt{-g^{(2)}}$ with $g^{(2)}$ is the determinant of this sector. In this section, we denote $a,b$ as two dimensional index unless they are mentioned explicitly.

    To find the new energy-momentum tensor $\bar{T}_{ab}$ under the conformal transformation
$\bar{g}_{ab} = \Omega^2 g_{ab}$, we consider that the form of the anomaly expressions 
do not change; i.e. 
\begin{equation}
\bar{T}\equiv \bar{g}^{ab}\bar{T}_{ab} = \frac{\bar{R}}{48\pi}~.
\label{2.02} 
\end{equation}
Next let us write the most general expression for the transformed energy-momentum tensor in terms of the old one and the 
conformal factor as
\begin{equation}
\bar{T}_{ab} = A T_{ab} + B \nabla_a\Omega\nabla_b\Omega + C g_{ab}\Box\Omega + 
D g_{ab}\nabla_c\Omega\nabla^c\Omega + E\nabla_a\nabla_b\Omega~,
\label{2.03} 
\end{equation}
where $A,B,C,D,E$ are unknown parameters which will be determined in the 
following way. The above structure is chosen such that $\bar{T}_{ab}$ reduces to $T_{ab}$ when $\Omega=1$ and correspondingly 
$A=1$.
It is known that under the conformal transformation the two dimensional Ricci scalar transforms as \cite{Carroll:2004st}
\begin{equation}
\bar{R} = \frac{R}{\Omega^2} - \frac{2}{\Omega^3}\Box\Omega + \frac{2}{\Omega^4}\nabla_a
\Omega\nabla^a\Omega~.
\label{2.04} 
\end{equation}
Substituting (\ref{2.03}) in the left hand side of (\ref{2.02}) and (\ref{2.04}) on the right hand side 
of the same equation and then using the first equation of (\ref{2.01}) we obtain
\begin{eqnarray}
&&A\Omega^{-2}\frac{R}{48\pi} + (B+2D)\frac{\nabla_a\Omega\nabla^a\Omega}{\Omega^2}
+(2C+E)\frac{\Box\Omega}{\Omega^2}
\nonumber\\
&& = \frac{1}{48\pi}\Big[\frac{R}{\Omega^2} - \frac{2}{\Omega^3}\Box\Omega 
+\frac{2}{\Omega^4}\nabla_a\Omega\nabla^a\Omega\Big]~.
\label{2.05} 
\end{eqnarray}
It tells that to satisfy the above equality we have following choices for the unknown parameters:
\begin{equation}
A=1; \,\,\ B+2D = \frac{2}{48\pi\Omega^2}; \,\,\ 2C+E=-\frac{2}{48\pi\Omega} ~.
\label{2.06} 
\end{equation}
Therefore, the transformation rule of stress-tensor in this case is
\begin{equation}
\bar{T}_{ab} = T_{ab} + B\nabla_a\Omega\nabla_b\Omega -  \Big(\frac{1}{48\pi\Omega}+\frac{E}{2}\Big)g_{ab}\Box\Omega 
+ \Big(\frac{1}{48\pi\Omega^2} - \frac{B}{2}\Big)g_{ab}\nabla_c\Omega\nabla^c\Omega + E\nabla_a\nabla_b\Omega~.
\label{2.07}
\end{equation}
It must be noted that in the above two parameters, $B$ and $E$, remain undetermined. Since for the present case, 
the conformal factor $\Omega =a$, which is given by (\ref{a}), has the dimension of squire length, $B$ must depends on 
conformal factor as $B\sim \Omega^{-2}$ while the dependence of $E$ on the conformal factor is $E\sim \Omega^{-1}$. 
Although the explicit expressions of them are hard to find within this analysis, we will see that
the nature of dependence on the conformal factor is sufficient for our later analysis \footnote{It may be interesting 
to see if use of second equation of (\ref{2.01}) in the conformal frame helps us to find explicit expressions of $B$ 
and $E$. But, as we will see, such a complication is unnecessary for the present analysis.}. For future purpose, 
remember that the bar quantities are for SD metric while the unbar ones are for Schwarzschild metric.

  In order to obtain the flux from the horizon of the SD metric, it is sufficient to find the 
component $\bar{T}^r_t$. Note that this is related to the components 
of energy-momentum tensor in null coordinates for the metric (\ref{2SD}) by the following relation:
\begin{equation}
\bar{T}^r_t = \bar{g}^{rr}\bar{T}_{rt} = a^{-2}F\bar{T}_{rt} = -\frac{1}{a^2}(\bar{T}_{uu} - \bar{T}_{vv})~.
\label{2.08}
\end{equation}
Now to evaluate the null components, (\ref{2.07}) will be used. It yields the relation of the components of 
stress-tensor between Schwarzschild metric and the SD metric in null coordinates as
\begin{eqnarray}
&&\bar{T}_{uu} = T_{uu} + B (\partial_u\Omega)^2 + E\nabla_u\nabla_u\Omega;
\nonumber
\\
&&\bar{T}_{vv} = T_{vv} + B (\partial_v\Omega)^2 + E\nabla_v\nabla_v\Omega~,
\label{2.09}
\end{eqnarray}
where the conformal factor is given by $\Omega = a$ which is Eq. (\ref{a}).
Therefore, we have 
\begin{equation}
\bar{T}^r_t = -\frac{1}{a^2}\Big[(T_{uu}-T_{vv}) + B\{(\partial_u\Omega)^2 - (\partial_v\Omega)^2\} 
+ E(\nabla_u\nabla_u\Omega-\nabla_v\nabla_v\Omega)\Big]~.
\label{2.09n1} 
\end{equation}
Next use of $\partial_u =1/2(\partial_t-F\partial_r)$ and $\partial_v=1/2(\partial_t+F\partial_r)$ yield,
\begin{eqnarray}
&&(\partial_u\Omega)^2 - (\partial_v\Omega)^2 = -F\partial_t\Omega\partial_r\Omega = -\frac{8ma}{r}~;
\nonumber
\\ 
&&\nabla_u\nabla_u\Omega-\nabla_v\nabla_v\Omega = -F\partial_t\partial_r\Omega + \frac{F'}{2}\partial_t\Omega 
=  F'a^{1/2}-\frac{4m}{r}~,
\label{2.09n2}
\end{eqnarray}
where in the above (\ref{a}) has been used.
The components of $T_{ab}$ in null coordinates in the case of Schwarzschild metric can be found out by solving the 
anomaly equations (\ref{2.01}).   
This has been done earlier. In stead of giving the details, let us just give the final expressions in null 
coordinates \cite{Banerjee:2008sn}:
\begin{eqnarray}
T_{uu} = -\frac{1}{96\pi}(FF'' - \frac{F'^2}{2})+C_{uu}~;\,\,\,\
T_{vv}=0~. 
\label{TCompo}
\end{eqnarray}
To fix the constant $C_{uu}$ we impose the Unruh vacuum condition which is the relevant vacuum for 
the Hawking radiation \cite{Banerjee:2008wq}. This is given by the vanishing of $T_{uu}$ near the horizon. Then (\ref{TCompo}) 
yields $C_{uu} = -\kappa^2/48\pi$. Therefore $T^r_t$, calculated in the limit $r\rightarrow\infty$, yields
\begin{equation}
T^r_t(r\rightarrow\infty) = -T_{uu} (r\rightarrow\infty) = \frac{\kappa^2}{48\pi}~,
\label{SCT} 
\end{equation}
where $\kappa = F'(r_H)/2 = 1/4m$ is the surface gravity of the Schwarzschild black hole. This is the Hawking 
flux radiated from the Schwarzschild horizon.

   Now substitution of (\ref{2.09n2}) and (\ref{TCompo}) in (\ref{2.09n1}) yields
\begin{equation}
\bar{T}^r_t = -\frac{1}{a^2}\Big[-\frac{1}{96\pi}(FF'' - \frac{F'^2}{2})+C_{uu} - B\frac{8ma}{r} + E(F'a^{1/2}-\frac{4m}{r})\Big]~.
\label{2.10} 
\end{equation}
As explained earlier, since $B\sim\Omega^{-2}$ and $E\sim\Omega^{-1}$ we see that $8Bma/r\sim 1/ar$, $EF'a^{1/2}\sim 1/(r^2a^{1/2})$ 
 and $4Em/r\sim 1/ar$. Therefore, in the large $r$ limit all the terms except the constant $C_{uu}$ are 
negligible. Hence the flux is given by 
\begin{equation}
\bar{T}^r_t = -a^{-2}C_{uu} = \frac{\kappa^2}{48\pi a^2}~.
\label{TRT2} 
\end{equation}
This is the energy flux emitted from the SD metric horizon with temperature
\begin{equation}
T = \frac{\kappa}{2\pi a} = \frac{1}{8\pi m a}~.
\label{new1} 
\end{equation}
So one can define an effective surface gravity as $\kappa_{\textrm{eff}} = \kappa/a$ such that 
$T = \kappa_{\textrm{eff}}/2\pi$. The similar result was obtained earlier in \cite{Saida:2007ru}.

At this point, let me give a comparative study between my result (\ref{new1}) and the earlier findings \cite{Sultana:2005tp,Faraoni:2007gq,Saida:2007ru}. As I mentioned in the introduction, there exists two forms of the temperature in literature: one is time independent \cite{Sultana:2005tp} and other is time dependent \cite{Saida:2007ru}. The authors of \cite{Sultana:2005tp} obtained the expression based on the existence of the conformal Killing vector for the SD metric which is null on the conformal Killing horizon. The formalism requires two conditions. The conformal factor $\Omega$ and the conformal Killing vector $\xi^a$ must satisfy $\Omega\rightarrow 1$ (or constant) and $\xi^a\xi_a\rightarrow -1$ at null infinity. This leads to a temperature which is constant everywhere on the conformal Killing horizon. Unfortunately, none of the above conditions are satisfied for the SD metric as in the present case $\xi^a\xi_a\rightarrow -a^2\neq-1$ (see (\ref{a})) at null infinity. To find the correct result it has been pointed in \cite{Saida:2007ru} that the physical temperature should be identified from the expression of the emission spectrum through the horizon. Here, I found it using this argument which leads to a time dependent form. This of course compatible with the following intuitive idea. The black hole is expanding and the temperature is inversely proportional to the physical mass. In addition, the mass must be related to the physical radius of the black hole which is changing with time. Hence it is expected that the temperature will become time dependent.      
  
  Now we will show that the temperature obtained above is actually related to the nonaffinity parameter,
calculated at the horizon, 
corresponding to a null vector. This is identical to the usual static case where the Killing vectors are 
null at the horizon. The nonaffinity parameter $\lambda$ is defined with respect to a null vector $l^a$ as,
\begin{equation}
l^a\nabla_al^b = \lambda l^b~.
\label{nonaffinity} 
\end{equation}  
Let us first start with the Schwarzschild metric. In this case we choose the null vector in null coordinates ($u,v$) as
 $l_a=(-F/2,0,0,0)$; $l^a = (0,1,0,0)$. Then Eq. (\ref{nonaffinity}) leads to $\lambda=F'(r)/2$ where the prime denotes 
the derivative with respect to the radial coordinate $r$. This is exactly the surface gravity $\kappa$ at the horizon $r=2m$. 
Next we shall evaluate the same in the conformally transformed metric. Consider an arbitrary vector satisfies $g_{ab}A^aA^b = \epsilon$, 
where $\epsilon$ is a fixed constant. Suppose the transformed vector also obeys the identical condition 
$\bar{g}_{ab}\bar{A}^a\bar{A}^b = \epsilon$. Then the transformation rule for the vector is $\bar{A}^a=\Omega^{-1}A^a$.
Therefore $l^a$ should transform 
as $l^a \rightarrow \Omega^{-1}l^a$. It immediately implies from Eq. (\ref{nonaffinity}) that
\begin{eqnarray}
\lambda\rightarrow \bar{\lambda} = \frac{\lambda}{\Omega} + \frac{l^a\nabla_a\Omega}{\Omega^2}~.
\label{lambdanew}
\end{eqnarray}
Now for the present case, $\Omega = a(t,r)$ since SD metric is conformal to Schwarzschild metric with the conformal factor $a^2(t,r)$ (See, Eq. (\ref{SDSC})) and $l^a\nabla_a\Omega = 
\partial_v\Omega$ where $\partial_v = 1/2(\partial_t + F(r)\partial_r)$. Therefore
\begin{equation}
\bar{\lambda} = \frac{F'}{2a} + \frac{1}{2} \Big(\frac{\dot{a}}{a^2} + \frac{Fa'}{a^2}\Big)~.
\label{lambda1} 
\end{equation}
The dot represents the derivative with respect to time coordinate $t$.
Next using the value of $a(t,r)$ from Eq. (\ref{a}) we obtain,
\begin{equation}
\frac{\dot{a}}{a^2} + \frac{Fa'}{a^2} = \frac{2}{a^{3/2}} \Big(1 + \frac{2m}{r}\Big)~,
\end{equation}
which, in the limit $r\rightarrow 2m$, vanishes more rapidly than the first term of Eq. (\ref{lambda1}). Hence keeping only the 
dominant term in our desire limit we find
$\bar{\lambda} = \kappa/a$, 
where $\kappa = F'(2m)/2$ is the surface gravity of the Schwarzschild black hole. It must be 
observed that $\bar{\lambda}$, calculated above, is precisely the $\kappa_{\textrm{eff}}$.

  Before concluding this section, couple of comments are as follows. At first, it must be noted that to evaluate the non-affinity parameter for SD case
the null vector is taken to be conformal to that of the Schwarzschild metric. Since any null vector for the seed metric is still a null vector for its 
conformally connected metric, one might wounder that the original vector is sufficient. But as the metric is conformally transformed, all the 
quantities should also be seen in the properly transformed basis. This is actually highlighted in the above as the temperatures, calculated by both ways, agree.  Finally, remember that the temperature was evaluated on the horizon location at $r=2m$. This I will show is the trapping horizon. 
The expansion parameter $\theta$ is defined as $\theta + \lambda = 
\nabla_al^a$ where $l^a$ is a null vector. Now since under the conformal transformation $\lambda$ is transformed as Eq. (\ref{lambdanew}), 
$\theta$ will transform by the relation
\begin{equation}
\bar{\theta} = \frac{\theta}{\Omega}+\frac{2}{\Omega^2}l^a\nabla_a\Omega~.
\label{thetatrans}
\end{equation}
This, for the present case, leads to $\bar{\theta} = F/ar + 2 (\dot{a}/a^2 + Fa'/a^2)$, which vanishes for $r=2m$.  
Therefore $r=2m$ is actually a trapping horizon and the above temperature is defined on this surface. 
In the next section, the thermodynamics of SD metric will be discussed on this horizon.

\section{\label{Smarr} Smarr formula and Law of black hole mechanics}
     The present section will mainly deal thermodynamics of the horizon of the SD metric. Here we will give the Smarr formula 
and show that the temperature obtained in the above is consistent with the first law of thermodynamics. The energy of the gravitating 
system will be considered as the Misner-Sharp energy \cite{Misner:1964je} which is ideal for a dynamical case, contrary to the Komar energy \cite{Komar:1958wp} defined with 
respect to a Killing vector.
 
   The relation between the Misner-Sharp energy of a conformally transformed metric with that of the seed metric which has spherical symmetry, 
is given by \cite{Faraoni:2014lsa}:
\begin{equation}
\bar{E} = \Omega E - \frac{r^3}{2\Omega}\nabla^a\Omega\nabla_a\Omega - r^2\nabla^a\Omega\nabla_a r~. 
\end{equation}
Use of it gives the energy of the SD metric as
\begin{equation}
\bar{E} = ma\Big[1+\frac{2r^3}{ma}(1+\frac{2m}{r})-\frac{4r}{a^{1/2}}\Big]~. 
\end{equation}
This has already been obtained earlier in \cite{Saida:2007ru,Faraoni:2014lsa}.
Now in the near horizon limit $r\rightarrow 2m$, the conformal factor $a$ diverges, and so keeping only the leading order terms we obtain the 
Misner-Sharp energy as $\bar{E}=ma$. The entropy of the SD metric is found out from the area law. It turns out to be
\begin{equation}
\bar{S} = \frac{\bar{A}}{4} = \frac{1}{4}\int_{\mathcal{H}}d\theta d\phi \sqrt{\sigma} =   \frac{1}{4}\int_{\mathcal{H}}d\theta d\phi a^2r^2\sin\theta = 4\pi m^2a^2~.
\end{equation}
Therefore at the horizon we obtain
\begin{equation}
 2\bar{S}T = ma=\bar{E}~.
\label{3.01}
\end{equation}
The similar relation was concluded earlier in the case of static or stationary black holes in general relativity \cite{Padmanabhan:2003pk,Banerjee:2010yd,Banerjee:2010ye}. 
There the energy was identified as 
the Komar conserved quantities. It has been shown in \cite{Banerjee:2010yd,Banerjee:2010ye} that such a general relation is actually the Smarr formula \cite{Smarr:1972kt}. (For more progress in this direction, see \cite{Padmanabhan:2013lpa}.) Here, from (\ref{3.01}) we find  
the Smarr formula for the SD metric in the following form:
\begin{equation}
\frac{ma}{2} = \frac{\bar{A}\kappa_{\textrm{eff}}}{8\pi}~, 
\end{equation}
where $\bar{A}$ is the horizon area. In addition it is also possible to obtain another relation which relates the differential form of the entropy and energy:
\begin{equation}
Td\bar{S} = d(ma)=d\bar{E}~. 
\end{equation}
This is the first law of black hole thermodynamics for the $r=2m$ horizon of SD spacetime. So far we observed that the temperature obtained in section \ref{temp} is consistent with the 
general structure of the Smarr formula and first law of black hole thermodynamics.

\section{\label{con}Conclusions}
   In this paper, the thermodynamics of the Sultana-Dyer (SD) black hole has been studied which is connected to the 
Schwarzschild black hole by a time dependent conformal factor. The temperature was obtained by using the expressions 
of the two dimensional gravitational anomalies. The horizon, considered here, was shown to be the apparent horizon. 
We showed that this temperature is connected to the Misner-sharp energy and entropy by a simple relation $\bar{E} = 
2\bar{S}T$. Interestingly, such a relation is identical to the general form of the Smarr formula for the stationary 
cases in GR. Using this, we obtained the Smarr formula for the SD metric. The first law of thermodynamics was also found. 
The analysis showed that our temperature is consistent with the thermodynamics of the horizon. This singles out one temperature 
among the two existing forms (one is time dependent and other is time independent) in literature if we insist the viability 
of the thermodynamic structure of the gravity in time dependent case.

   Let us now summarise what we achieved so far. In addition to the thermodynamics of the SD metric, the analysis shows a simple way of 
finding the thermodynamical entities for a metric which is conformally connected to a static seed metric. In the present case the SD metric is 
time dependent. So this paper also exhibits, so far as I know, a first application of the anomaly method to study Hawking radiation for an evolving case.
Further we mention that, the temperature, obtained here, must be crossed checked by evaluating the radiation spectrum within other methods, like 
tunneling method (For a review discussing dynamical cases, see \cite{Vanzo:2011wq}), solving Wheeler-DeWitt equations \cite{Banerjee:2010xi}. 
The computations in this direction are in progress.

    Finally, it must be pointed out that for the metric (\ref{SD1}), the energy conditions are satisfied if $\eta<r(r+2m)/2m$. In this region, both the time-like and null dusts fall towards the black hole. So the accretion of the dusts increases the mass. On the other hand, for $\eta>r(r+2m)/2m$, the source fields in the Einstein's equations are unphysical. In this region, the dusts become superluminal near the horizon. For $r=2m$, the energy conditions are obeyed when $\eta<4m$. So at late time the particles near the horizon will be superluminal. For details, please look at the refs. \cite{Sultana:2005tp,Saida:2007ru}. Although it has such weakness, since its global structure is similar to that of a cosmological black hole, it would be interesting to study different aspects of this simple example. That will help us to understand more realistic situations. Moreover, the presence of this unpleasant superluminal feature is may be due to the fact that the present model is far from the realistic one. It is expected that the exact solutions in both the GR and other alternative theories of gravity will not contain these problems.

\vskip 9mm
\noindent 
{\bf Acknowledgements}\\
I thank Valerio Faraoni for useful discussions.
The research of the author is supported by a Lady Davis Fellowship at Hebrew University, by the
I-CORE Program of the Planning and Budgeting Committee and the Israel Science Foundation
(grant No. 1937/12), as well as by the Israel Science Foundation personal grant No. 24/12.

\begin{appendix}
\section*{\bf{Appendix}}
\section*{\label{Appendix} Dimensional reduction}
\begin{figure}[h!]
    \centering
    \includegraphics[width=0.5\textwidth]{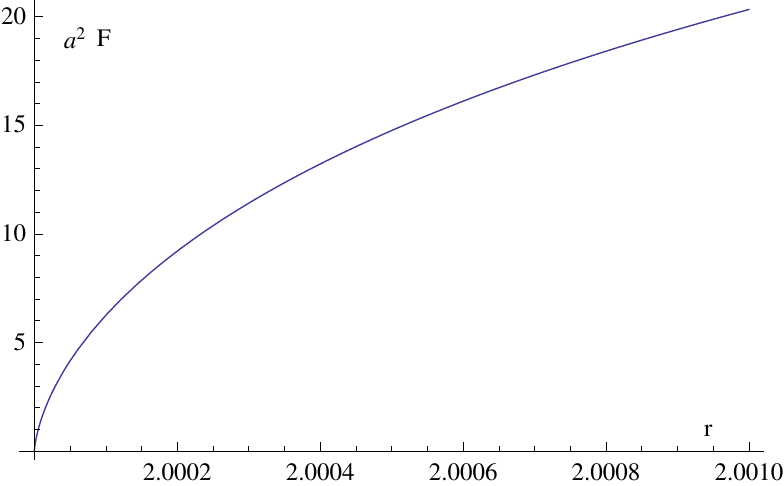}
    \caption{$a^2F$ Vs. $r$ plot for $m=1=t$.}
\label{fig1}
\end{figure}

\begin{figure}[h!]
    \centering
    \includegraphics[width=0.5\textwidth]{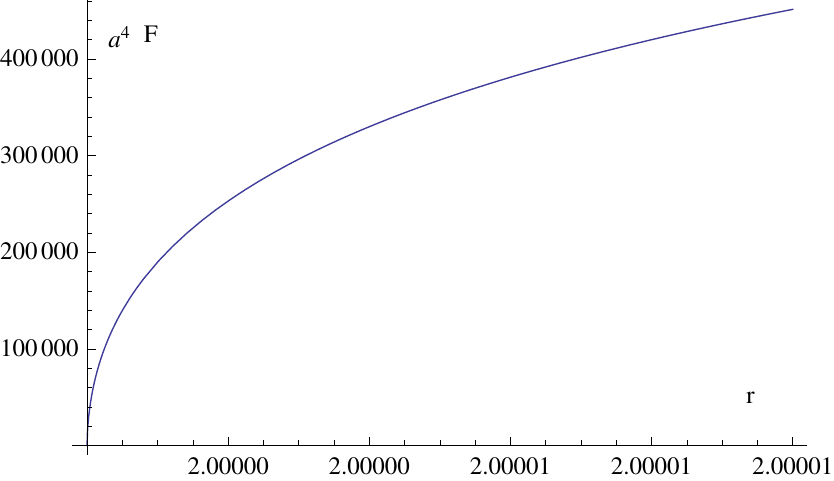}
    \caption{$a^4F$ Vs. $r$ plot for $m=1=t$.}
\label{fig2}
\end{figure}

\renewcommand{\theequation}{A.\arabic{equation}}
\setcounter{equation}{0}  
     Consider the massive Klein-Gordon equation $\Box\Phi+M^2\Phi=0$ under the SD metric (\ref{SDSC}). 
Expanding this equation we find
\begin{eqnarray}
&&-\frac{r^2\sin\theta}{F}\partial_t(a^2\partial_t\Phi) + \sin\theta\partial_r(a^2r^2F\partial_r\Phi)
+a^2\cos\theta\partial_\theta\Phi+a^2\sin\theta\partial^2_\theta\Phi
\nonumber
\\
&&+\frac{a^2}{\sin\theta}\partial^2_\phi\Phi
+M^2a^4r^2\sin\theta\Phi = 0~. 
\end{eqnarray}
In the tortoise coordinate $dr_* = dr/F$, the above reduces to
\begin{eqnarray}
&&-\frac{r^2\sin\theta}{F}\partial_t(a^2\partial_t\Phi) + \frac{\sin\theta}{F}\partial_{r_*}(a^2r^2\partial_{r_*}\Phi)
+a^2\cos\theta\partial_\theta\Phi+a^2\sin\theta\partial^2_\theta\Phi
\nonumber
\\
&&+\frac{a^2}{\sin\theta}\partial^2_\phi\Phi
+M^2a^4r^2\sin\theta\Phi = 0~.  
\end{eqnarray}
Multiplying both sides by $F/(r^2\sin\theta)$ we obtain
\begin{equation}
-\partial_t(a^2\partial_t\Phi) + \frac{1}{r^2}\partial_{r_*}(a^2r^2\partial_{r_*}\Phi)
-\frac{a^2F}{r^2}L^2\Phi+M^2a^4F\Phi = 0~, 
\end{equation}
where the angular operator $L^2 = -\partial^2_\theta-\cot\theta\partial_\theta-(1/\sin^2\theta)\partial^2_\phi$.
Now expanding the field in spherical harmonics as $\Phi(t,r_*,\theta,\phi) = \displaystyle\sum_{l,m}
\Psi_{lm}(t,r_*)Y_{lm}(\theta,\phi)$ and then substituting in the above one gets
\begin{equation}
\displaystyle\sum_{l,m}Y_{lm}\Big[-\partial_t(a^2\partial_t\Psi_{lm})+\frac{1}{r^2}\partial_{r_*}(a^2r^2
\partial_{r_*}\Phi_{lm})-\frac{a^2F}{r^2}l(l+1)\Psi_{lm}+M^2a^4F\Psi_{lm}\Big]=0 
\end{equation}
where $L^2Y_{lm}(\theta,\phi) = l(l+1)Y_{lm}(\theta,\phi)$ has been used. Next multiplying both sides by 
$\Phi^*(t,r_*,\theta,\phi) = \displaystyle\sum_{l,m}\Psi_{lm}^*(t,r_*)Y_{lm}^*(\theta,\phi)$ and then integrating 
over the angular coordinates we obtain,
\begin{equation}
\displaystyle\sum_{l,m}\Psi_{lm}^*\Big[-\partial_t(a^2\partial_t) +\frac{1}{r^2}\partial_{r_*}(a^2r^2\partial_{r_*})
 - \frac{a^2F}{r^2}l(l+1)+M^2a^4F\Big]\Psi_{lm} = 0~.
\end{equation}
To obtain the final expression the orthogonality relation $\int d\theta d\phi\sin\theta Y^*_{l'm'}Y_{lm}
=\delta_{ll'}\delta_{mm'}$ was used. In the near horizon limit, keeping only the dominating terms, the above 
reduces to
\begin{eqnarray}
\displaystyle\sum_{l,m}\Psi_{lm}^*\Big[-\partial_t(a^2\partial_t) +\frac{1}{r^2}\partial_{r_*}(a^2r^2\partial_{r_*})
\Big]\Psi_{lm} = 0~. 
\end{eqnarray}
Returning back to the original radial coordinate we find in the near horizon limited
\begin{equation}
\displaystyle\sum_{l,m}\Psi_{lm}^*\Big[-\frac{1}{F}\partial_t(a^2\partial_t) +\partial_{r}(a^2F\partial_{r})
\Big]\Psi_{lm} = 0~. 
\label{app12}
\end{equation}
Since $\Psi_{lm}$'s are independent modes, each mode will satisfy the following equation
\begin{equation}
\Big[-\frac{1}{F}\partial_t(a^2\partial_t) +\partial_{r}(a^2F\partial_{r})
\Big]\Psi_{lm} = 0~. 
\label{app}
\end{equation}
This is the Klein-Gordon equation of motion for each mode under the two dimensional metric
\begin{equation}
ds^2_{\textrm{eff}} = a^2(t,r)\Big[-F(r)dt^2+\frac{dr^2}{F(r)}\Big]~. 
\end{equation}
To arrive (\ref{app12}) the terms proportional to $a^2F$ and $a^4F$ have been discard near $r=2m$. Note that although 
$a$ diverges while $F=0$ in this limit, the graphical representations of them (Figure \ref{fig1} and Figure \ref{fig2}) confirm 
that these terms tend to zero 
very near to the horizon. 
   
\end{appendix}


\begin{thebibliography}{99}
\bibitem{Sultana:2005tp} 
  J.~Sultana and C.~C.~Dyer,
  Gen.\ Rel.\ Grav.\  {\bf 37}, 1347 (2005).

\bibitem{Faraoni:2007gq} 
  V.~Faraoni,
  Phys.\ Rev.\ D {\bf 76}, 104042 (2007)
  [arXiv:0710.2122 [gr-qc]].

\bibitem{Saida:2007ru} 
  H.~Saida, T.~Harada and H.~Maeda,
  Class.\ Quant.\ Grav.\  {\bf 24}, 4711 (2007)
  [arXiv:0705.4012 [gr-qc]].

\bibitem{Hawking:1974sw} 
  S.~W.~Hawking,
  Commun.\ Math.\ Phys.\  {\bf 43}, 199 (1975)
  [Erratum-ibid.\  {\bf 46}, 206 (1976)].

\bibitem{Misner:1964je} 
  C.~W.~Misner and D.~H.~Sharp,
  Phys.\ Rev.\  {\bf 136}, B571 (1964).

\bibitem{Padmanabhan:2003pk} 
  T.~Padmanabhan,
  Class.\ Quant.\ Grav.\  {\bf 21}, 4485 (2004)
  [gr-qc/0308070].

\bibitem{Banerjee:2010yd} 
  R.~Banerjee and B.~R.~Majhi,
  Phys.\ Rev.\ D {\bf 81}, 124006 (2010)
  [arXiv:1003.2312 [gr-qc]].\\
  B.~R.~Majhi,
  J.\ Phys.\ Conf.\ Ser.\  {\bf 405}, 012020 (2012)
  [arXiv:1209.5876 [gr-qc]].

\bibitem{Banerjee:2010ye} 
  R.~Banerjee, B.~R.~Majhi, S.~K.~Modak and S.~Samanta,
  Phys.\ Rev.\ D {\bf 82}, 124002 (2010)
  [arXiv:1007.5204 [gr-qc]].

\bibitem{Smarr:1972kt} 
  L.~Smarr,
  Phys.\ Rev.\ Lett.\  {\bf 30}, 71 (1973)
  [Erratum-ibid.\  {\bf 30}, 521 (1973)].

\bibitem{Faraoni:2013aba} 
  V.~Faraoni,
  Galaxies {\bf 1}, no. 3, 114 (2013)
  [arXiv:1309.4915 [gr-qc]].

\bibitem{Majhi:2011yi} 
  B.~R.~Majhi,
  ``Quantum Tunneling in Black Holes,'' (Thesis)
  arXiv:1110.6008 [gr-qc].

\bibitem{AlvarezGaume:1983ig} 
  L.~Alvarez-Gaume and E.~Witten,
  Nucl.\ Phys.\ B {\bf 234}, 269 (1984).

\bibitem{Polyakov:1981rd} 
  A.~M.~Polyakov,
  Phys.\ Lett.\ B {\bf 103}, 207 (1981).

\bibitem{Leutwyler:1984nd}
  H.~Leutwyler,
  Phys.\ Lett.\  B {\bf 153}, 65 (1985)
  [Erratum-ibid.\  {\bf 155B}, 469 (1985)].

\bibitem{Christensen:1977jc} 
  S.~M.~Christensen and S.~A.~Fulling,
  Phys.\ Rev.\ D {\bf 15}, 2088 (1977).

\bibitem{Robinson:2005pd} 
  S.~P.~Robinson and F.~Wilczek,
  Phys.\ Rev.\ Lett.\  {\bf 95}, 011303 (2005)
  [gr-qc/0502074].

\bibitem{Banerjee:2007qs} 
  R.~Banerjee and S.~Kulkarni,
  Phys.\ Rev.\ D {\bf 77}, 024018 (2008)
  [arXiv:0707.2449 [hep-th]].

\bibitem{Carroll:2004st} 
  S.~M.~Carroll,
  ``Spacetime and geometry: An introduction to general relativity,''
  San Francisco, USA: Addison-Wesley (2004) 513 p

\bibitem{Banerjee:2008sn} 
  R.~Banerjee and B.~R.~Majhi,
  Phys.\ Rev.\ D {\bf 79}, 064024 (2009)
  [arXiv:0812.0497 [hep-th]].

\bibitem{Banerjee:2008wq} 
  R.~Banerjee and S.~Kulkarni,
  Phys.\ Rev.\ D {\bf 79}, 084035 (2009)
  [arXiv:0810.5683 [hep-th]].

\bibitem{Komar:1958wp} 
  A.~Komar,
  Phys.\ Rev.\  {\bf 113}, 934 (1959).

\bibitem{Faraoni:2014lsa} 
  V.~Faraoni and V.~Vitagliano,
  Phys.\ Rev.\ D {\bf 89}, 064015 (2014) 
  [arXiv:1401.1189 [gr-qc]].

\bibitem{Padmanabhan:2013lpa} 
  T.~Padmanabhan,
  Gen.\ Rel.\ Grav.\  {\bf 46}, 1673 (2014)
  [arXiv:1312.3253 [gr-qc]].

\bibitem{Vanzo:2011wq} 
  L.~Vanzo, G.~Acquaviva and R.~Di Criscienzo,
  Class.\ Quant.\ Grav.\  {\bf 28}, 183001 (2011)
  [arXiv:1106.4153 [gr-qc]].

\bibitem{Banerjee:2010xi}  
  C.~Kiefer, J.~Muller-Hill, T.~P.~Singh and C.~Vaz,
  Phys.\ Rev.\ D {\bf 75}, 124010 (2007)
  [gr-qc/0703008].\\
  R.~Banerjee, C.~Kiefer and B.~R.~Majhi,
  Phys.\ Rev.\ D {\bf 82}, 044013 (2010)
  [arXiv:1005.2264 [gr-qc]].
\end{thebibliography}
\end{document}